\documentclass[nofootinbib,pra,twocolumn]{revtex4-1}

\newcommand{\beq}{\begin{eqnarray}}
\newcommand{\eeq}{\end{eqnarray}}

\newcommand{\ann}{{\text a}} 
\newcommand{\tts}{t_{\text s}} 
\newcommand{\btau}{\mathbf{\tau}}

\newcommand{\comment}[1]{}

\usepackage{graphicx,amsmath}
\usepackage{epstopdf}
\usepackage{hyperref}

\begin{document}

\title{Unraveling Quantum Annealers using Classical Hardness}
\author{Victor Martin-Mayor}
\affiliation{Departamento de F\'isica Te\'orica I, Universidad Complutense, 28040 Madrid, Spain}
\affiliation{Instituto de Biocomputaci\'on y F\'isica de Sistemas Complejos (BIFI), Zaragoza, Spain}
\author{Itay Hen}
\affiliation{Information Sciences Institute, University of Southern California, Marina del Rey, California 90292, USA}

\date{\today}

\begin{abstract}
Recent advances in quantum technology have led to the development and
manufacturing of experimental programmable quantum annealing
optimizers that contain hundreds of quantum bits. These optimizers,
named `D-Wave' chips, promise to solve practical optimization problems
potentially faster than conventional `classical' computers.  Attempts
to quantify the quantum nature of these chips have been met with both
excitement and skepticism but have also brought up numerous
fundamental questions pertaining to the distinguishability of quantum
annealers from their classical thermal counterparts. Here, we propose
a general method aimed at answering these, and apply it to
experimentally study the D-Wave chip. Inspired by
spin-glass theory, we generate optimization problems with a wide
spectrum of `classical hardness', which we also define. By
investigating the chip's response to classical hardness, we
surprisingly find that the chip's performance scales unfavorably as
compared to several analogous classical algorithms. We detect, quantify and discuss
purely classical effects that possibly mask the
quantum behavior of the chip.
\end{abstract}

\maketitle

\section{Introduction}
Interest in quantum computing originates in the
potential of quantum computers to solve certain computational problems
much faster than is possible classically, due to the unique properties
of Quantum Mechanics~\cite{shor:94,grover:97}. The implications of
having at our disposal reliable quantum computing devices are
obviously tremendous.
The actual implementation of quantum computing devices is however
hindered by many challenging difficulties, the most prominent being
the control or removal of quantum
decoherence~\cite{schlosshauer:04}. In the past few years, quantum
technology has matured to the point where limited, task-specific,
non-universal quantum devices such as quantum communication systems,
quantum random number generators and quantum simulators, are being
built, possessing capabilities that exceed those of corresponding
classical computers.

Recently, a programmable quantum annealing machine, known as the
D-Wave chip~\cite{johnson:11}, has been built whose goal is to
minimize the cost functions of classically-hard optimization problems
presumably by adiabatically quenching quantum fluctuations. If
found useful, the chip could be regarded as a prototype for
general-purpose quantum optimizers, due to the broad range of hard theoretical and practical problems that may be encoded on it. 

The capabilities, performance and underlying physical mechanism
driving the D-Wave chip have generated fair amounts of curiosity,
interest, debate and controversy within the Quantum Computing
community and beyond, as to the true nature of the device and its
potential to exhibit clear ``quantum signatures''.  While some studies
have concluded that the behavior of the chip is consistent with
quantum open-system Lindbladian dynamics~\cite{albash:12} or
indirectly observed entanglement~\cite{lanting:14}, other studies
contesting these~\cite{smolin:13,shin:14} pointed to the existence of
simple, purely classical, models capable of exhibiting the main
characteristics of the chip.

Nonetheless, the debate around the quantum nature of the chip has raised several fundamental questions pertaining to the manner in which quantum devices should be characterized in the absence of clear practical ``signatures'' such as (quantum) speedups~\cite{ronnow:14,boixo:14}. Since quantum annealers are meant to utilize an altogether different mechanism for solving optimization problems than traditional classical devices, methods for quantifying this difference are expected to serve as important theoretical tools while also having vast practical implications. 

Here, we propose a method that partly solves the above question by
providing a technique to characterize, or measure, the
``classicality'' of quantum annealers. This is done by studying the
algorithmic performance of quantum annealers on sets of optimization
problems possessing quantifiable, varying degrees of ``thermal'' or
``classical'' hardness, which we also define for this purpose. To
illustrate the potential of the proposed technique, we apply it to the
experimental quantum annealing optimizer, the D-Wave Two (DW2) chip.

We observe several distinctive phenomena that reveal a strong
correlation between the performance of the chip and classical
hardness: i) The D-Wave chip's typical time-to-solution ($\tts$) as a
function of classical hardness scales differently, in fact worse, than
that of thermal classical algorithms. Specifically, we find that
the chip does very poorly on problem instances exhibiting a phenomenon known as
``temperature chaos''. ii) Fluctuations in success probability between programming cycles become larger with increasing hardness, pointing to the fact that encoding errors become more pronounced and destructive with increasing hardness. iii) The success probabilities obtained from harder instances are affected more than easy instances by changes in the duration of the anneals. 

Analyzing the above findings, we identify two major probable causes for the
chip's observed ``sub-classical'' performance, namely i) that its temperature may not be low enough, and  ii) that encoding errors become more pronounced with increasing hardness.
We further offer experiments and simulations
designed to detect and subsequently rectify these so as to enhance
the chip's capabilities.  

\section{ Classical Hardness, temperature chaos and parallel tempering}

In order to study the
manner in which the performance of quantum annealers correlates with
`classical hardness', it is important to first accurately establish
the meaning of \emph{classical hardness\/}. For that purpose, we refer
to spin-glass theory~\cite{young:98}, which deals with {\it spin
  glasses}--- disordered, frustrated spin systems that may be viewed
as prototypical classically-hard (also called NP-hard) optimization
problems, that are so challenging that specialized hardware has been
built to simulate them~\cite{janus:08,janus:09,janus:14}.

Currently, the (classical) method of choice to study general
spin-glass problems is Parallel Tempering (PT, also known as `exchange
Monte Carlo')~\cite{hukushima:96,marinari:98b}. PT is a refinement of
the celebrated yet somewhat outdated Simulated Annealing
algorithm~\cite{kirkpatrick:83}, that finds optimal assignments (i.e.,
the ground-state configurations) of given discrete-variable cost functions. 
It is therefore only natural to make use of  the performance
of PT to characterize and quantify classical hardness. 

\begin{figure}
\begin{center}
\includegraphics[angle=270,width=\columnwidth]{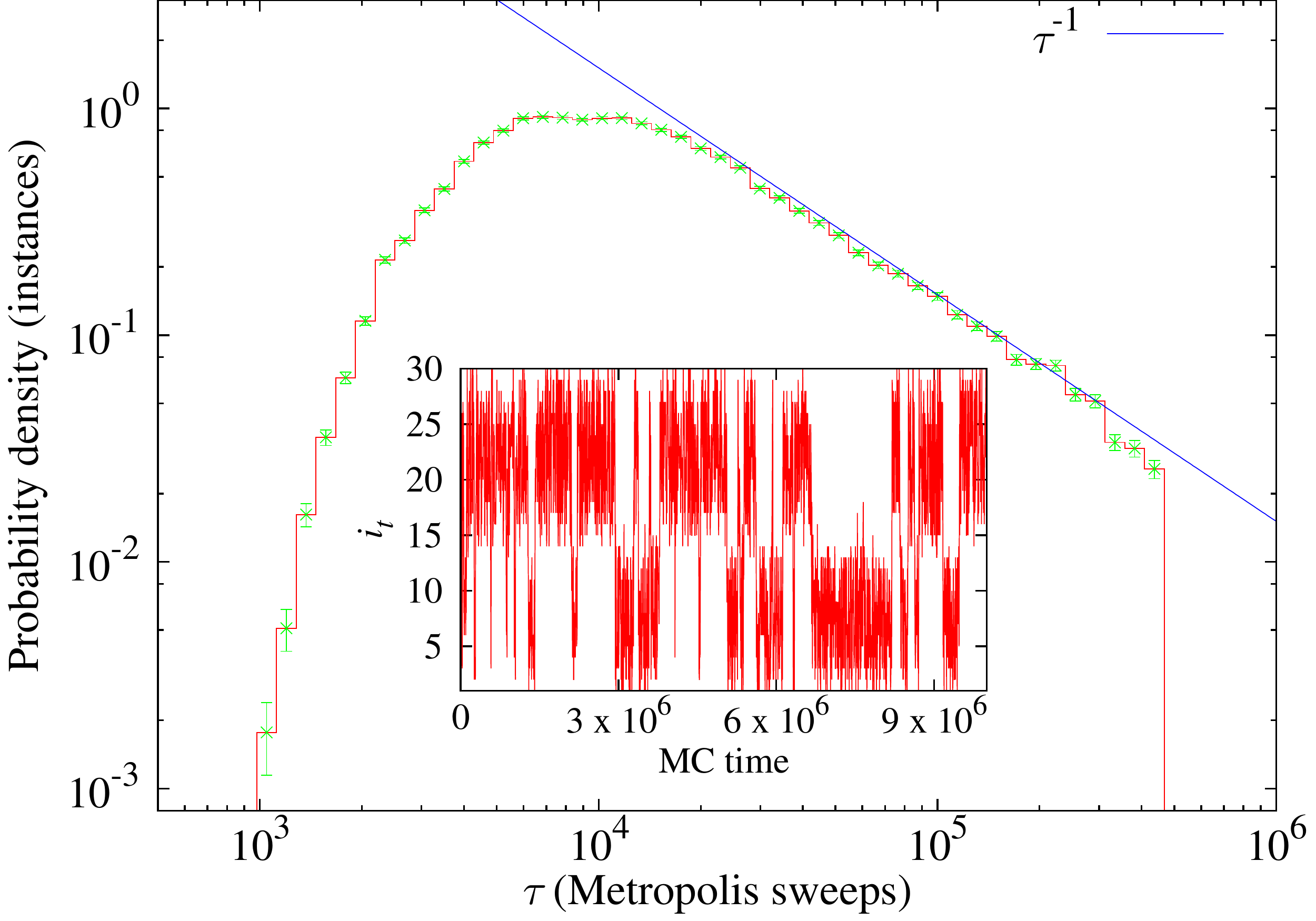}
\caption{{\bf Probability distribution of mixing times $\tau$ over
    random $J=\pm1$ `Chimera' instances as extracted from the PT
    random-walk on the temperature grid.}  The solid line is a linear fit to
  the tail of the distribution, implying the existence of rare
  instances with very long mixing times (note that a $\tau^{-1}$ tail is not strictly
  integrable, hence this specific power law decay should only be regarded as a finite-size approximation).  {\bf
    Inset:} Example of a temperature random walk for an instance with
  $\tau\approx 1.8\times 10^5$. Considering one of $N_T=30$ copies of the
  system, at any given Monte Carlo time $t$, the copy's temperature is
  $T_{i_t}$. In this example, the replica has visited each temperature
  several times, pointing to the fact that the simulation time is
  longer than the mixing time $\tau$. }
\label{fig:tau_distrib}
\vspace{-0.7cm}
\end{center}
\end{figure}

In PT simulations, one considers $N_T$ copies of an $N$-spin system at
temperatures $T_1<T_2<\ldots< T_{N_T}$, where each copy undergoes Metropolis
spin-flip updates independently of other copies. In addition, copies
with neighboring temperatures regularly attempt to swap their
temperatures with probabilities that satisfy detailed balance~\cite{sokal:97}. In this
way, each copy performs a temperature random-walk (see inset of Fig~\ref{fig:tau_distrib}). At high temperatures, free-energy
barriers are easily overcome, allowing for a global exploration of
configuration space. At lower temperatures on the other hand, the local
minima are explored in more detail.  A `healthy' PT simulation
requires an unimpeded temperature flow: The total length of the
simulation should  be longer than the temperature `mixing
time' $\tau$~\cite{fernandez:09b,janus:10}. The time $\tau$ may be thought of as  the
average time it takes each copy to fully traverse the temperature
mesh, indicating equilibration of the simulation. Therefore, instances with
large $\tau$ are harder to equilibrate, which motivates our definition
of the mixing time $\tau$ as the \emph{classical hardness} of a given instance.

\begin{figure}
\begin{center}
\includegraphics[angle=270,width=\columnwidth]{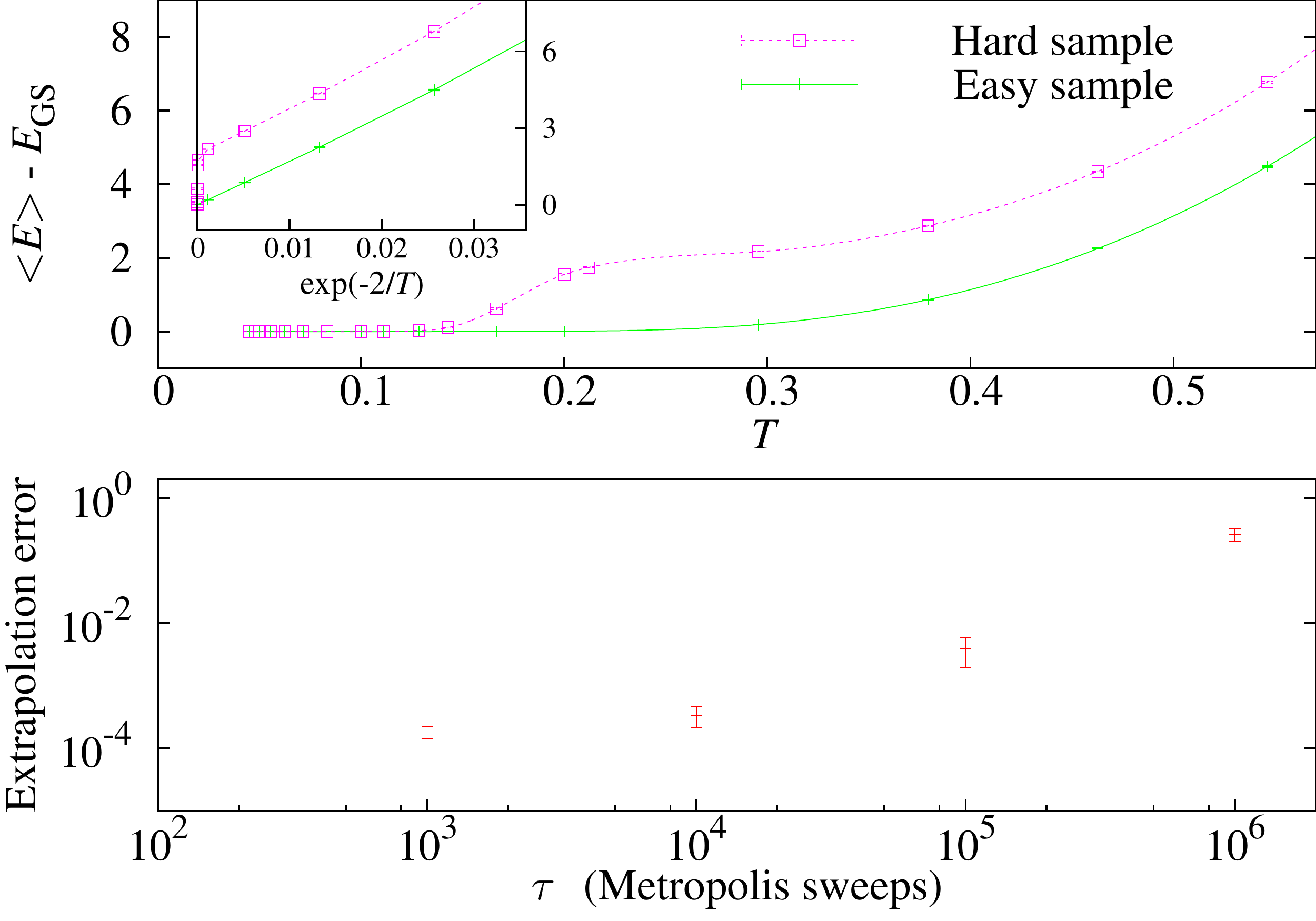}
\caption{{\bf Top: Energy above the ground-state energy
  as a function of temperature for two randomly chosen instances.}
  The mixing times of the two instances are $\tau \approx 10^3$ (``easy'') and $\tau \approx
  10^6$ (``hard''). Unlike the easy sample, the hard sample exhibits ``temperature chaos'':
  Upon lowering the temperature, the energy decreases at first in a gradual manner, however at $T\approx 0.2$ there is a sudden drop indicating that a different set of
  minimizing configurations has been visited.  {\bf Inset:} Main panel's
  data vs.  $\text{exp}(-\Delta/T)$ (where $\Delta=2$ is the excitation gap). A linear behavior is expected if the system can be described as a gas of non-interacting
excitations over a local energy minimum. For the easy sample,
this local minimum is a ground state. On the other hand, for instances
displaying TC above their chaos temperature, the local-minimum 
energy is higher than the ground-state's.  {\bf Bottom: $\tau$-dependence of the median systematic error  (for each $\tau$-generation) of a
  $T\to 0$ extrapolation of the total energy.} For each instance, we extrapolated the $T=0.2,0.3$ data linearly in $\text{exp}(-\Delta/T)$ and compared this extrapolation with the actual ground-state energy.}
\label{fig:tempChaos}
\vspace{-0.7cm}
\end{center}
\end{figure}

Despite the popularity of PT, it has also become apparent that not all
the spin-glass problems can be efficiently solved by the
algorithm~\cite{janus:10,fernandez:13}. The reason is a
phenomenon that has become known as {\emph{Temperature Chaos}}
(TC). TC~\cite{mckay:82,bray:87b,banavar:87,kondor:89,kondor:93,billoire:00,rizzo:01,mulet:01,billoire:02,krzakala:02,rizzo:03,parisi:10,sasaki:05,katzgraber:07,fernandez:13,billoire:14}
consists of a sequence of first-order phase transitions that a given
spin-glass instance 
experiences upon lowering its
temperature, whereby the dominant configurations minimizing the free
energy above the critical temperatures are vastly different than those
below them (Fig.~\ref{fig:tempChaos} depicts such a phase that is `rounded' due to the finite size of the system). 
A given instance may experience
zero, one or more transitions at random temperatures, making the
study of TC excruciatingly
difficult~\cite{sasaki:05,katzgraber:07,fernandez:13,billoire:14}. 
Such TC transitions hinder the PT
temperature flow, significantly prolonging the mixing time $\tau$. 
In practice~\cite{fernandez:13}, it is found that for small systems 
the large majority of the instances do not suffer any TC transitions
and are `easy' (i.e., they are characterized by short mixing times). However, for a minor fraction of
them, $\tau$ turns out to be inordinately large. Moreover, the larger the system is, the larger the fraction of long-$\tau$ samples becomes. In the large $N$ limit, these are the short-$\tau$ samples that become exponentially rare in $N$~\cite{rizzo:03,fernandez:13}. 
This provides further motivation for studying TC instances of optimization problems on moderately small experimental devices  (even if they are rare).

\section{Temperature chaos and quantum annealers}

With the advent of quantum annealers, which presumably offer non-thermal
mechanisms for finding ground states, it has become only natural to ask
whether quantum annealers can be used to solve `TC-ridden' optimization
problems faster than classical techniques such as PT. In this context, the
question of how the performance of quantum annealers depends on the `classical
hardness' becomes of fundamental interest: If indeed quantum annealers exploit
quantum phenomena such as tunneling to traverse energy barriers, one may hope
that they will not be as sensitive to the thermal hardness (as defined above)
of the optimization problems they solve. As we shall see next, having a
practical definition for classical hardness allows us to address the above
questions directly.

To illustrate this, in what follows we apply the ideas introduced above to the
DW2 quantum annealing optimizer in order to infer its degree of
`classicality'.  We accomplish this by first generating an ensemble of
instances that are directly embeddable on the DW2 `Chimera' architecture [the
  reader is referred to Appendix~\ref{sect:DW2} for
  a detailed description of the Chimera lattice and the D-wave chip and its
  properties]. The chip on which we perform our study is an array of 512
superconducting flux qubits of which only 476 are functional, operating at a
temperature of $\sim 15$mK. The DW2 chip is designed to solve a very specific
type of problems, namely, Ising-type optimization problems, by adiabatically
transitioning the system Hamiltonian from an initial transverse-field
Hamiltonian to a final classical programmable cost function of a typical spin
glass.  The latter is given by the Ising Hamiltonian: \beq
\label{eq:H} H_{\text{Ising}}=\sum_{\langle ij\rangle} J_{ij} s_i
s_j + \sum_i h_i s_i \,. 
\eeq 
The Ising spins, $s_i=\pm 1$ are the variables to be optimized over,
and the sets $\{J_{ij}\}$ and $\{h_i\}$ are programmable
parameters of the cost function.  Here, $\langle ij\rangle$
denotes a sum over all the active edges of the Chimera graph.  For simplicity, we conduct our study on 
randomly-generated problem instances with $h_i=0$ and random, equiprobable $J_{ij}=\pm J$ couplings (in our energy units $J=1$).

Initially, it is not clear whether the task of finding thermally-hard
instances on the Chimera is feasible. While on the one hand TC has
been observed in spin-glasses on the square lattice~\cite{thomas:11},
which has the same spatial dimension, $D=2$, as the
Chimera~\cite{katzgraber:14}, it has also been found that typical
Chimera-embeddable instances are easy to
solve~\cite{ronnow:14,boixo:14,katzgraber:14}.  As discussed above,
system size plays a significant role in this context, as an $N\sim
512$-spin Chimera may simply be too small to have instances exhibiting
TC.\footnote{For instance, on the square lattice one needs to reach $N\sim512^2\approx 2.5\times 10^5$
  spins for TC to be the rule rather than the exception~\cite{thomas:11}.}  Taking a brute-force approach to resolve this issue, we
generated $\sim80,000$ random problem instances (each characterized by a different set of $\{J_{ij}\}$),  analyzing each one by
running them on a state-of-the-art PT algorithm until equilibration was
reached (Appendix~\ref{sect:PT}).  This allowed for the calculation of the instances'
classical hardness, namely their temperature mixing times $\tau$ (for
more details, see Appendix~\ref{sect:Methods}, below). The resulting distribution of
$\tau$ over the instances is shown in Fig.~\ref{fig:tau_distrib}.
While most instances equilibrate rather quickly (after some $10^4$ MC
steps), we find that the distribution has a `tail' of hard samples
with $\tau > 10^6$ revealing that hard instances, although rare, do
exist (we estimate that 2 samples in $10^4$ have $\tau>10^7$).

To study the DW2 chip, we grouped together instances with similar
classical hardness, i.e., similar mixing times, $10^k \leq \tau \leq 3
\cdot 10^k$ for $k=3,4,5,6$ and $7$. For each such `generation' of
$\tau$, we randomly picked 100 representative instances for running on
the chip (only 14 instances with $k=7$ were found).  As
a convergence test of PT on the selected instances, we verified that the
ground-state energies reached by PT are the true ones by means of an exact solver.

At the purely classical level, we found, as
anticipated~\cite{fernandez:13}, that classically hard instances
differ from easy instances from a thermodynamic point of view as well.
Specifically, large $\tau$ instances were found to exhibit sharp
changes in the average energy at random critical temperatures,
consistently with the occurrence of TC (see
Fig.~\ref{fig:tempChaos}). For such instances, the true ground states
are present during the simulations only below the TC critical
temperatures.  As the inset shows, the larger $\tau$ is, the lower
these critical temperatures typically are. Furthermore,
classically-hard instances were found to differ from easier ones in
terms of their energy landscape: While for easy instances
minimally-excited states typically reside only a few spin flips away
from ground state configurations, for classically hard instances, this
is not the case (see Appendix~\ref{sect:PT}).

\section{Classicality of the ``D-Wave Two'' chip}

Having sorted and analyzed the randomly-generated instances, we turned to
experimentally test the performance of the D-Wave chip on these (for details
see Appendix~\ref{sect:Methods} and~\ref{sect:ta}). Our experiments consisted of programming the
chip to solve each of the 414 instances over a dense mesh of annealing times
in the available range of $t_{\ann} \in [20\mu s,20ms]$. The number of
attempts, or anneals, that each instance was run for each choice of $t_{\ann}$
ranged between $10^5$ and $10^8$. By calculating the success probability of
the annealer for each instance and annealing time, a typical time-to-solution
$\tts$ was obtained for each hardness-group, or `$\tau$-generation' (see Appendix~\ref{sect:Methods}). Interestingly, we found that for easy samples ($\tau=10^3$)
the success probability depends only marginally on $t_{\ann}$, pointing to the
annealer reaching its asymptotic performance on these. As instances become
harder, the sensitivity of success probability to $t_{\ann}$ increases
significantly.  Nonetheless, for all hardness groups, the typical $\tts$ is
found to be shortest at the minimally-allowed annealing time of
$t_{\ann}=20\mu$s (see Appendix~\ref{sect:ta} for a more detailed discussion).

\begin{figure}
\begin{center}
\includegraphics[width=\columnwidth]{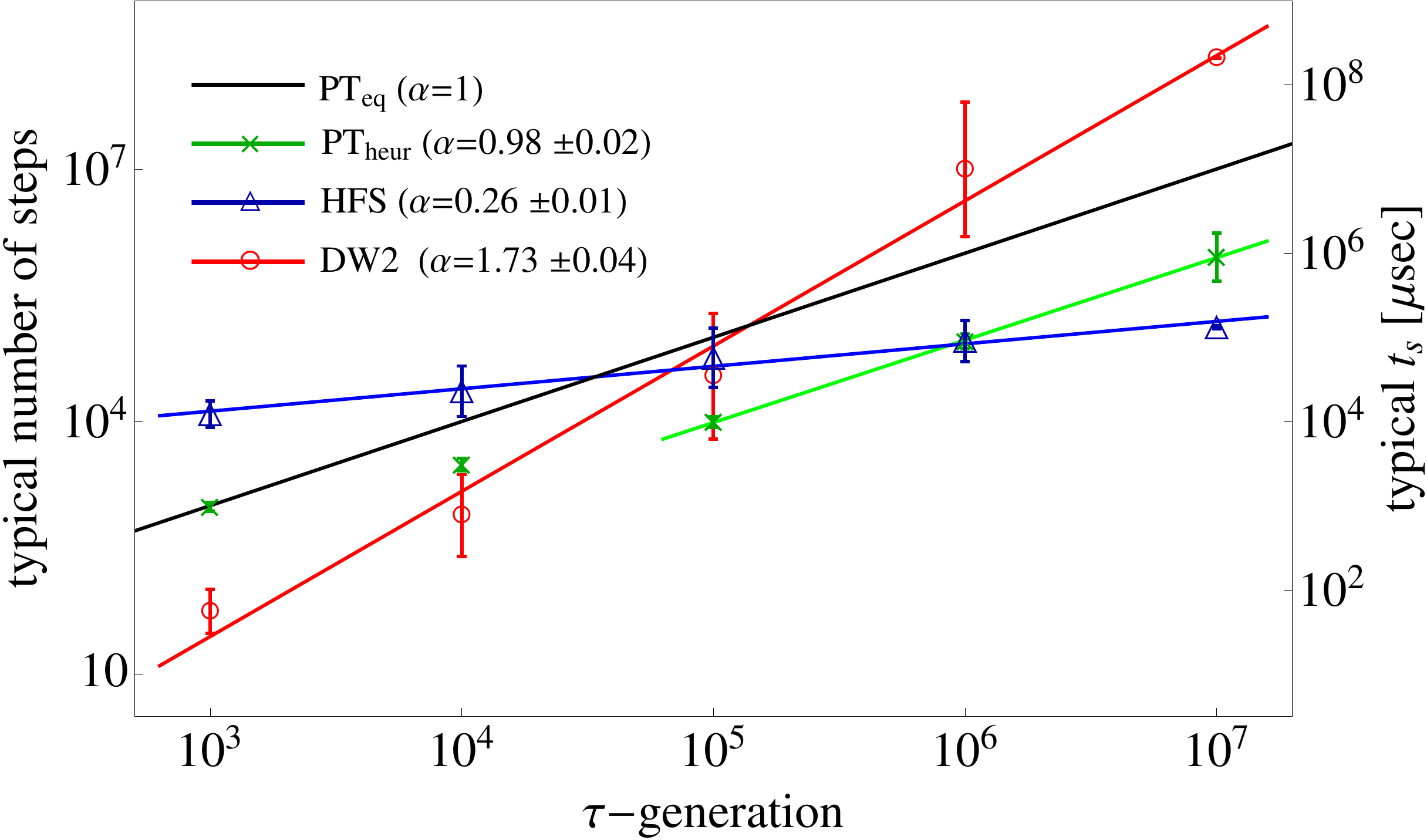}
\caption{{\bf Dependence of typical time to solution $\tts$ of the examined optimization algorithms on mixing time $\tau$, the classical hardness parameter.} 
Classical thermal algorithms scale linearly with 
$\tau$. Here, ${\text{PT}}_{\text{eq}}$ denotes time to equilibration which by definition scales linearly with $\tau$, and ${\text{PT}}_{\text{heur}}$ denotes PT functioning as a heuristic solver in which case the time to solution is the number of Monte-Carlo steps to first encounter of a minimizing configuration.
The $\tts$ of the classical non-thermal HFS algorithm (measured in $\mu$s) scales as 
$\tau^{\alpha_{\text{HFS}}}$, with  $\alpha_{\text{HFS}}\approx 0.3$. The $\tts$ for the DW2 chip (measured in $\mu$s) scales as $\tau^{\alpha_{\text{DW2}}}$
with $\alpha_{\text{DW2}}\approx1.73$ (we note the missing error bars on the $10^7$ DW2 data point, which is a result of insufficient statistics). 
}
\label{fig:tau_scaling}
\vspace{-0.7cm}
\end{center}
\end{figure}

The main results of our investigation are summarized in
Fig.~\ref{fig:tau_scaling} depicting the typical time to solution
$\tts$ of the DW2 chip (averaged over instances of same hardness
groups, see Appendix~\ref{sect:Methods}) as a function of classical hardness, or
`$\tau$-generation'.  As is clear from the figure, the performance of
the chip was found to correlate strongly with the `thermal hardness'
parameter, indicating the significant role thermal hardness plays in the annealing
process.  Interestingly, the chip's response was found to be affected
by thermal hardness even more than PT, i.e., more strongly than the
classical thermal response: While for PT the time-to-solution scales
as $\tts \sim \tau$, the scaling of the D-Wave chip was found to scale
as $\tts^{\text{DW2}} \sim \tau^{\alpha_{\text{DW2}}}$, with
$\alpha_{\text{DW2}} \approx 1.73$.  This scaling is rather
surprising given that for quantum annealers to perform better than
classical ones, one would expect these to be less susceptible to
thermal hardness, not more. Nonetheless, it is clear that the notion
of classical hardness is very relevant to the D-wave chip.

To complete the picture, we have also tested our instances on the
Hamze-de Freitas-Selby (HFS) algorithm~\cite{hamze:04,selby:14}, which
is the fastest classical algorithm to date for Chimera-type
instances. Even though the HFS algorithm is a `non-thermal'
algorithm (i.e., it does not make use of a temperature parameter),
we have found the concept of classical hardness to be very relevant
here as well. For the HFS algorithm, we find a scaling of
$\tts^{\text{HFS}} \sim \tau^{\alpha_{\text{HFS}}}$ with
${\alpha_{\text{HFS}}}\approx 0.3$, implying that the algorithm is significantly
less susceptible to thermal hardness than PT.\footnote{It
  is worth pointing out that the typical runtime for the HFS
  algorithm on the hardest, $\tau=10^7$, group problems was found to be $\sim
  0.5$s on an Intel Xeon CPU E5462 @ 2.80GHz, which to our knowledge
  makes these the hardest known Chimera-type instances to date.}

\section{Analysis of findings} The above somewhat less-than-favorable
performance of the experimental D-wave chip on thermally-hard problems
is not necessarily a manifestation of the intrinsic limitations of
quantum annealing, i.e., it does not necessarily imply that the
`quantum landscape' of the tested problems is harder to traverse than
the classical one (although this may sometimes be the
case~\cite{hen:11, farhi:12}).  A careful analysis of the results
suggests in fact at least two different more probable
`classical' causes for the chip's performance.

First, as already discussed above and succinctly captured in
Fig.~\ref{fig:tempChaos}, temperature is expected to play a key role in DW2
success on instances exhibiting TC. This is because for these, the ground
state configurations minimize the free energy only below the lowest critical
TC temperature.  Even though the working temperature of the DW2 chip is rather
low, namely $\sim 15$mK, the crucial figure of merit is the ratio of coupling
to temperature $T/J$ [recall Eq.~(\ref{eq:H})]. Although the nominal value for
the chip is $T/J \approx 0.1$, any inhomogeneity of the temperature across the
chip may render the ratio higher~\cite{kingPrivate}, possibly driving it above
typical TC critical temperatures.\footnote{We refer to the \emph{physical}
  temperature of the chip. However, non-equilibrium systems (e.g., supercooled
  liquids or glasses) can be characterized by \emph{two}
  temperatures~\cite{cugliandolo:97}: On the one hand, the physical
  temperature $T$ which rules fast degrees of freedom that equilibrate. On the other
  hand, the `effective' temperature refers to slow degrees of freedom that
  remain out of equilibrium. We are currently investigating whether or not the
  DW2 chip can analogously be characterized by two such temperatures.}

Another possible cause for the above scaling may be due to the analog nature
of the chip.  The programming of the coupling parameters $J_{ij}$ and
magnetic fields $h_i$ is prone to statistical and systematic errors (also referred to as intrinsic control errors, or ICE).
The couplings actually encoded in DW2 are $J_{ij}=\pm J + R$, where
$R \sim \mathcal{N}(0,\delta J)$ is a random error ($\delta J\approx 0.05 J$,
according to the chip's manufacturer). Unfortunately, even tiny changes in coupling values 
are known to potentially change the ground-state configurations of spin
glasses in a dramatic manner~\cite{nifle:92,ney-nifle:98,krzakala:05,katzgraber:07}.  We
refer to this effect as `coupling chaos' (or $J$-chaos, for
short). For an $N$-bit system, $J$-chaos seems to become significant
for $\delta J_{\textrm{crit}} \sim |J|/
N^{a}$. Empirically~\cite{krzakala:05,katzgraber:07} $a\approx 1/D$,
$D$ being the spatial dimension of the system.  Note,
however, that these estimates refer only to typical instances and small
$N$ whereas the assessment of the effects of $J$-chaos on
thermally-hard instances remains an important open problem for
classical spin glasses.

\begin{figure}
\begin{center}
\includegraphics[angle=270,width=\columnwidth]{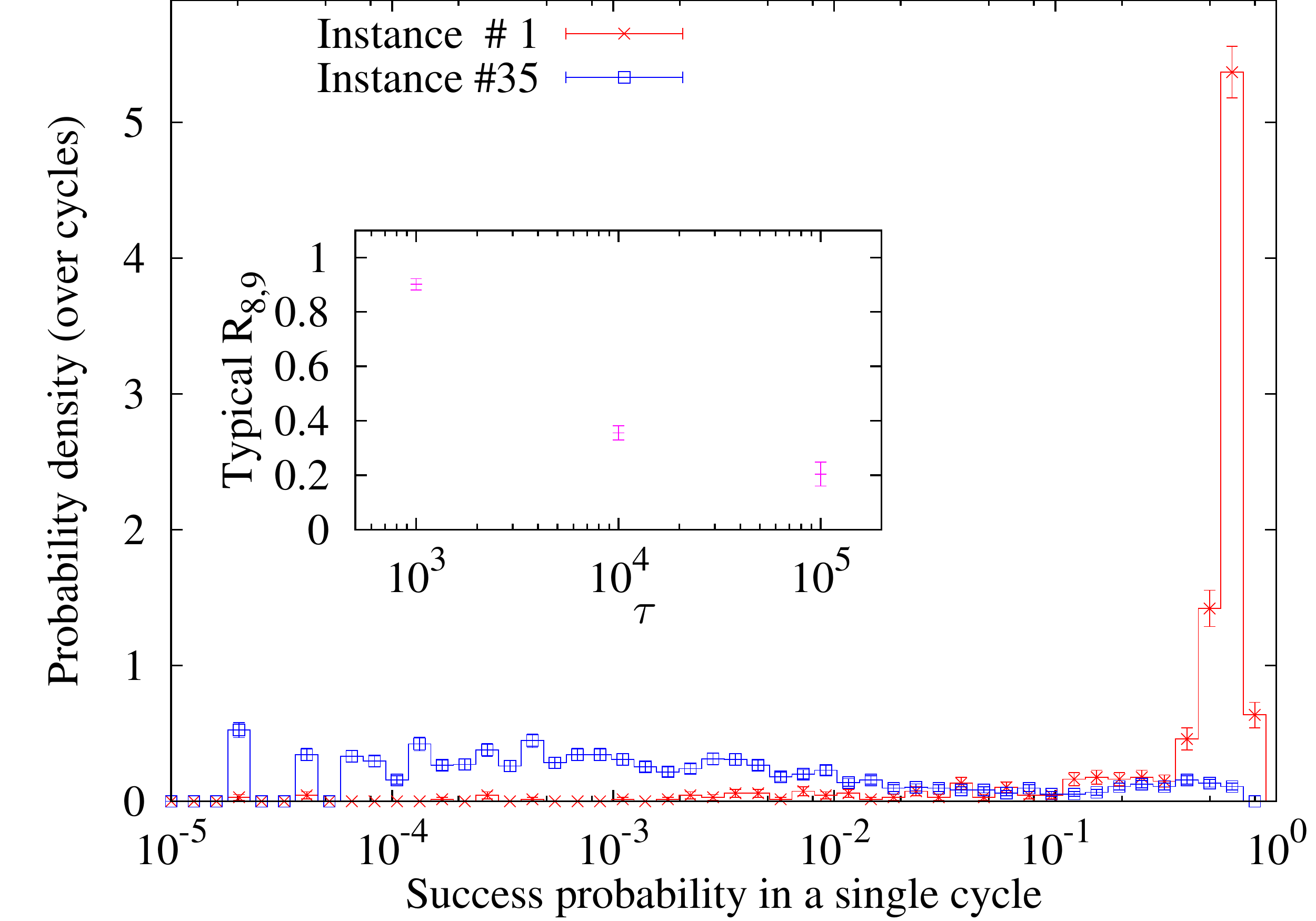}
\caption{{\bf Empirical evidence for the absence/presence of
    `$J$-chaos'.}  Probability density of success probability of a single cycle,
  $p$, over different
  programming cycles. The probability densities are plotted here for two easy ($\tau=10^3$) instances
 (here, $t_\ann=20\mu$s and
  the number of anneals per programming cycle is $X=49500$, see Appendix~\ref{sect:Methods}. Instance \#1 (662 cycles) is typical in this $\tau$-generation
  with success probability $p\sim 1$ in the majority of the
  programming cycles. On the other hand, instance \#35 (1624 cycles) suffers from
  strong $J$-chaos: Even though the probability of finding $p\sim 0.1$
  in some of the programming cycles is not negligible, most cycles are
  significantly less successful, e.g., the median $p$ is
  $I_{q=0.5}=6.7(5)\times 10^{-4}$.  {\bf Inset:}
  Typical ratio of the $80$th percentile
  probability to the $90$th percentile probability, namely
  $R_{8,9}=I_{q=0.8}/I_{q=0.9}$ (see text) as a function of $\tau$
  (for $\tau\geq 10^6$ and $10^7$, $R_{8,9}$ was not computed due to
  extremely low success probabilities). Smaller ratios indicate larger fluctuations in success probabilities.}
\label{fig:J_Chaos}
\vspace{-0.7cm}
\end{center}
\end{figure}

Here, we empirically quantify the effects of $J$-chaos by taking advantage
of the many programming cycles and gauge choices each instance has
been annealed with (typically between 200 and 2000). Calculating a success probability $p$ for each cycle, we compute the probability distribution of
$p$ over different cycles for each instance~\cite{fernandez:13,billoire:14}. We find that while
for some instances $p$ is essentially insensitive to programming
errors, for other instances (even within the same thermal hardness
group), $p$ varies significantly, spanning several orders of
magnitude. This is illustrated in Fig.~\ref{fig:J_Chaos} which
presents some results based on a straightforward percentile analysis
of these distributions. For instance \#1 in the figure,
the $80$th percentile probability is $I_{q=0.8}(p)=0.669(3)$, whereas
the probability at the $90$th percentile is
$I_{q=0.9}(p)=0.698(4)$. Hence the ratio
$R_{8,9}=I_{q=0.8}(p)/I_{q=0.9}(p)$ is close to one. Conversely, for instance \#35, the values are $I_{q=0.8}(p)=0.008(1)$,
$I_{q=0.9}(p)=0.07(2)$ and the ratio is $R_{8,9}=0.12$, i.e., the
success probability drops by an order of magnitude. The inset of
Fig.~\ref{fig:J_Chaos} shows the typical ratio $R_{8,9}$ as a function
of classical hardness, demonstrating the strong correlation between
thermal hardness and the devastating effects of $J$-chaos caused by ICE, namely that
the larger $\tau$ is, the more probable it is to find instances for
which $p$ varies wildly between programming cycles.  \\ 

\section{Discussion} 

We have devised a method for quantifying the `classicality' of quantum annealers
by studying their performance on sets of instances characterized by different degrees of thermal hardness, which we have defined for that purpose as the mixing 
(or equilibration) time $\tau$ of classical thermal algorithms (namely, PT) on these. 
We find that the 2D-like Chimera architecture used in the D-Wave chips does give rise to thermally very-hard, albeit rare, instances. Specifically, we have found samples that exhibit temperature chaos, and as a such have very long mixing times, i.e., they are classically exceptionally hard to solve.

Applying our method to an experimental quantum annealing optimizer,  the DW2 chip,
 we have found that its performance is more susceptible to changes in thermal hardness than classical algorithms.
This is in contrast with the performance of the best-known state-of-the-art classical solver on Chimera graphs, the `non-thermal' HFS algorithm,
which scales (unsurprisingly) better. Our results are not meant to suggest
that the DW2 chip is not a quantum annealer, but rather that
its quantum properties may be significantly masked by much-undesired classical effects.

We have identified and quantified two probable causes for the observed
behavior: A possibly too high temperature, or more probably, $J$-chaos, the random errors stemming from the
digital-to-analog conversion in the programming of the coupling
parameters.  One may hope that the scaling of current DW2 chips would
significantly improve if one or both of the above issues are
resolved. Clearly, lowering the temperature of the chip and/or
reducing the error involved in the programming of its parameters are
both technologically very ambitious goals, in which case error correcting techniques may prove very useful~\cite{pudenz:14}. 
We believe that 
quantum Monte Carlo simulations of the device will be
instrumental in the understanding of the roles
that temperature and magnitude of programming errors play in the performance of the chip (and of  its classical counterparts).
In turn, this will help sharpening the most pressing technological
challenges facing the fabrication of these and other future quantum
optimizing devices, paving the way to obtaining a long-awaited
insights as  to the difference between quantum and classical hardness
in the context of optimization. We are currently pursuing these
approaches. 

\section*{Acknowledgments}

We thank Luis Antonio Fern\'andez and David Yllanes for
providing us with their analysis program for the PT correlation
function. We also thank Marco Baity-Jesi for helping us to prepare the
figures.

We are indebted to Mohammad Amin, Luis Antonio Fern\'andez, Enzo Marinari, Denis
Navarro, Giorgio Parisi, Federico Ricci-Tersenghi and Juan Jes\'us
Ruiz-Lorenzo for discussions. 

We thank Luis Antonio Fern\'andez, Daniel Lidar, Felipe LLanes-Estrada, David
Yllanes and Peter Young for their reading of a preliminary version of the
manuscript.

We thank D-Wave Systems Inc. for granting us access to the chip. 

We acknowledge the use of algorithms and source code for a
classic solver, devised and written by Alex Selby, available for
public usage at~\url{https://github.com/alex1770/QUBO-Chimera}.

IH acknowledges support by ARO grant number W911NF-12-1-0523. VMM was
supported by MINECO (Spain) through research contract N$^{o}$
FIS2012-35719-C02.

\appendix
\section{Methods}\label{sect:Methods}

{\bf Computation of the mixing time $\btau$}.---
Because we follow Ref.~\cite{janus:10}, we just briefly summarize here the main steps of the procedure. 
Considering one of the $N_T$ system copies in the PT
simulation, let us denote the temperature of copy $i$ at Monte Carlo time $t$ by 
$T_{i_t}$, where $1\leq i_t\leq N_T$
(see inset of Fig.~\ref{fig:tau_distrib}).  At equilibrium, the probability
distribution for $i_t$ is uniform (namely, $1/N_T$) hence the exact
expectation value of $i_t$ is $\langle i_t\rangle=(N_T+1)/2$. From the general theory of Markov Chain Monte
Carlo~\cite{sokal:97}, it follows that the
equilibrium time-correlation function may be written as a sum of exponentially decaying terms:
\beq
C_{\text{PT}}(s)&=&\langle i_t i_{t+s}\rangle -\frac{(N_T+1)^2}{4}\\\nonumber
&=&\sum_{n} a_n \text{e}^{-s/\tau_{n}}\quad (\tau_1>\tau_2>\ldots)\,.
\eeq
The mixing time $\tau$ is the largest `eigen-time'
$\tau_1$. We compute numerically the correlation function
$C_{\text{PT}}(s)$ and fit it to the decay of \emph{two} exponential
functions (so we extract the dominant time scale $\tau$ and a
sub-leading time scale). The procedure is described in full in Ref.~\cite{janus:10}.
\\
\phantom{A}
\\
{\bf D-wave data acquisition and analysis}.--- 
\\
\phantom{A}
\\
\emph{Data acquisition:}
\\
In what follows we briefly summarize the steps of the experimental setup and data acquisition for the anneals performed on the 414 randomly-generated instances in the various thermal-hardness groups. 
\begin{enumerate}
\item 
The $J_{ij}$ couplings of each of the 414 instances have been encoded onto the D-Wave chip using many different choices of annealing times in the allowed range of
$20\mu$s$\leq t_\ann\leq$ $20m$s.
\item 
For each instance and each choice of $t_\ann$ the following process has been repeated hundreds to thousands of times:
\begin{enumerate}
\item \label{ia}
First, a random `gauge' has been chosen for the instance. A gauge transformation does not change the properties of the optimization problem but has some effect on the performance of the chip which follows from the imperfections of the device that break the gauge symmetry. The different gauges are applied by transforming the original instance $h_i\rightarrow \eta_i h_i$, $J_{ij}\rightarrow \eta_i \eta_j J_{ij}$, to the original cost function
  Eq.~\eqref{eq:H}. The above gauge transformations correspond to 
the change $s_i\rightarrow \eta_i s_i$ in configuration spin values. Here, the $N$ gauge parameters $\eta_i=\pm 1$ were chosen randomly.
\item 
The chip was then programmed with the gauge-transformed instance (inevitably adding programming bias errors, as mentioned in the
  main text).
  \item 
The instance was then solved, or annealed, $X$ times within the programming cycle/with the chosen gauge. We chose $X \approx 1 \textrm{sec}/t_\ann$, the maximally allowed amount.
\item 
After the $X$ anneals were performed, the number of successes $Y$, i.e., the number of times a minimizing configuration had  been found, was recorded. The probability of
  success for the instance, for that particular gauge/programming cycle and annealing time $t_\ann$ was then
  estimated as $p=Y/X$. Note, that in cases where the probability of success is of the order of $1/X$, the probability $p$ will be a rather noisy estimate (see data analysis, below, for a procedure to mitigate this problem).
\item If the prefixed number of cycles for the current instance and $t_{\ann}$ has not been reached, return to (\ref{ia}) and choose a new gauge.
\end{enumerate}
\end{enumerate}

\phantom{A}

\emph{Data analysis and time-to-solution estimates:}

The analysis of the data acquisition process described above proceeded as follows. 
\begin{enumerate}
\item  
For each instance and each annealing time, the total number of hits $Y_\text{tot}=\sum_i Y_i$ was calculated, where $i$ sums over all the gauge/programming cycles. Denoting $X_\text{tot}=\sum_i X_i$ as the total number of annealing
attempts, the probability of success for any particular instance and anneal time was then calculated as $P=Y_{\text{tot}}/X_{\text{tot}}$.
\item 
The above probability was then converted into an average time-to-solution $\tts$ for that
instance and $t_\ann$ according to $\tts=t_\ann/P$, where the special case  of $P=0$ designates an estimate of an infinite $\tts$, where in practice 
the true probability lies below the resolution threshold of $1/X_{\text{tot}}$.
\item 
A typical runtime for a hardness group was then obtained by taking the median over all minimal $\tts$ values of all the instances in the group. 

\end{enumerate}

\section{Instance generation and analysis}\label{sect:PT}

\subsection{The Parallel Tempering simulations}

We briefly outline the technical details of the Parallel Tempering (PT) simulations and subsequent analysis performed on the randomly generated instances.
The reader is referred to Ref.~\cite{janus:10} for further details.

To run the simulations, we employed multi-spin
coding~\cite{newman:99}, a simple yet efficient form of
parallel computation, that allows the simultaneous simulations of a large
number of problem instances. The name \emph{multi-spin} follows from
the fact that our dynamic variables are binary, $s_i=\pm 1$, and thus can be coded in a single bit.  Hence, one may code the
$i$-th spin corresponding to $M$ independent instances in a single
$M$-bit computer word. Using streaming extensions, $M$ can be as large
as $M=256$, nowadays.\footnote{Simulations were run on a standard Intel(R)
  Xeon(R) processors (E5-2690 0 @ 2.90GHz).} The advantages of
simulating in parallel 256 instances are obvious, if we want to study
a \emph{huge} number of problems in our quest for these rare instances
displaying Temperature Chaos. In fact, we have simulated a total of
$N_S=303\times 256=77568$ problem instances.

The temperature grid of the PT simulations consisted of $N_T=30$ temperatures. Temperatures with
indices $i=13,14,\ldots, N_T$ were evenly distributed in the range
$0.21 \leq T_i \leq T_\text{max}=1.632$, while lower temperatures in
the range $T_\mathrm{min}=0.045 \leq T_i \leq 0.2$ (indices
$i=1,2,\ldots,12$) have been added in order to detect temperature
chaos effects.\footnote{In our simulations, the minimal temperature value $T_\mathrm{min}=0.045$ corresponds effectively to zero temperature.  
This follows from the 
minimal energy gap of our instances being $\Delta=2$ together with our use of a 64-bit (pseudo) random-number generator~\cite{fernandez:09c}.
Setting $T_\mathrm{min}$ such that it obeys $2^{64}\times
\text{e}^{-2/T_\mathrm{min}}<1$, our Metropolis
simulations at $T_\mathrm{min}$ effectively become equivalent to $T=0$
simulations.} Ergodicity was maintained by the
temperature-swap part of the PT.

For each instance, we ran four independent simulations (i.e., four
replicas) per temperature, where an elementary Monte Carlo step
consisted of 10 full lattice Metropolis sweeps, followed by a PT
temperature swap attempt. Since the Chimera lattice is bipartite, Metropolis sweeps have been carried out on alternating partitions
at each step.

The calculation of the mixing time $\tau$ was conducted in three
stages, or rounds (see also Ref.~\cite{janus:10}).  In the first stage, all
$303\times 256$ instances were simulated for a total of $10^6$
elementary Monte Carlo steps (i.e., each system was simulated for
$10^7$ full-lattice Metropolis sweep). At the end of each round the
mixing time $\tau$ was computed (measured in units of full-lattice Metropolis
sweeps). As can also be read off Fig.~1 of the main text, the
first-round of simulations was adequate for the equilibration of most instances, namely, for
problems of $\tau$-generations $10^3,10^4$ and $10^5$.  The
second round of simulations was set to last 10 times longer, consisting of $10^7$ elementary Monte Carlo steps, or $10^8$ full-lattice
Metropolis sweeps per system. These longer simulations were reserved
only to the $1024$ hardest instances corresponding to $\tau$-generations of
$10^6$ or larger.\footnote{The criterion for selecting the hardest instances to qualify for subsequent rounds of simulations was done by assigning 
a figure of merit for each instance based on the performance of its $4\times 30=120$ system copies. For each of the
copies, the fraction of Monte Carlo time spent at the higher-temperatures regions, namely $T_i$ with $i=16,17,\ldots, N_T$, was calculated
where the figure of merit was chosen to be the smallest of these fractions, which was used as an indication for 
how trapped the instance is in the low-temperatures region.}  The 256 worse-scoring instances of these were further simulated for $10^8$ elementary Monte Carlo
steps (or $10^9$ full-lattice Metropolis sweeps per system). This
simulation lasted 10 days on our fastest CPU [Intel(R) Core(TM)
i7-4770K CPU @ 3.50GHz]. From it, we obtained 14 extremely hard
problem instances, belonging to $\tau$-generation $10^7$. 

\subsection{Spin-overlap analysis}

Here, we briefly extend the analysis of the energy landscapes
of thermally-hard problems, initiated in Fig.~2 of the main
text. To do so, we make use of the notion of `spin-overlap', which plays a major role in any spin glass investigation (see, e.g., Ref.~\cite{young:98}).
The spin overlap $q$ between two $N$-spin configurations
$\{s_i^{a}\}$ and $\{s_i^{b}\}$ is defined as
\begin{equation}\label{eq:overlap}
q=1- 2\frac{N_{a,b}}{N}\,,
\end{equation}
where $N_{a,b}$ is the Hamming distance between the two configurations, i.e., the number of spins by which the two configurations \emph{differ\/}.
For two identical configurations we get $N_{a,b}=0$ and
$q=1$ while for two configurations differing by a global spin
flip (i.e., $s_i^{a}=-s_i^{b}$ for all $i$) we have $N_{a,b}=N$ and
$q=-1$. Since the instances we study all have $h_i=0$, their Hamiltonians possess a global bit-flip symmetry. For any configuration $\{s_i^{a}\}$, its spin-reversed configuration $\{s_i^{b}\}=-\{s_i^{a}\}$ has the same energy. 
We have therefore chosen to consider the more informative measure of the absolute value of the overlap,
$|q|$. In this case, the maximum meaningful Hamming distance between
any two configurations is $N/2$.

\begin{figure}
\begin{center}
\includegraphics[angle=270,width=\columnwidth]{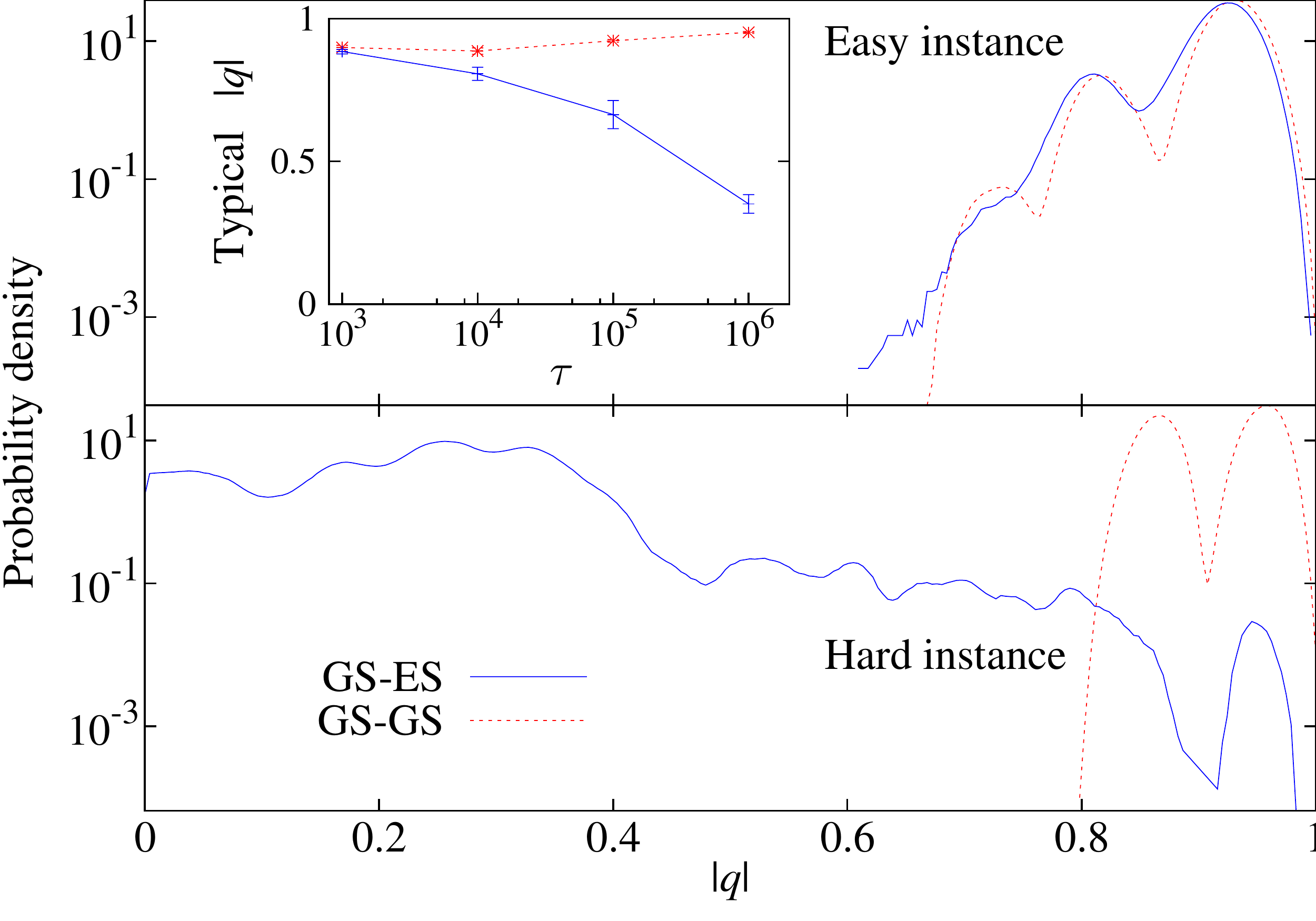}
\caption{{\bf Probability distribution of the spin overlap $|q|$
    between ground states and minimally excited states (GS-ES, solid blue curves).} For comparison, we also show the overlap between
  different ground states (GS-GS, dashed red curves). The distributions shown
  were computed for the same instances considered in Fig.~2 of main
  text, with mixing time generations $\tau \approx 10^3$ ({\bf top})
  and $\tau \approx 10^6$ ({\bf bottom}). {\bf Inset: Dependence of
    the overlap between ground states and minimally excited states on
    $\tau$.} Typical median GS-ES overlap $|q|$ averaged over each
  hardness group as a function of $\tau$-generation (blue points, the
  lines are to guide the eye).The red points are GS-GS typical
  overlaps shown for comparison.}
\label{fig:Pq}
\vspace{-0.7cm}
\end{center}
\end{figure}

We utilized spin overlaps to investigate the energy landscapes of the various instances in the
following way. During the PT simulations, 12000 spin
configurations of each instance were stored on disk, corresponding to 100 evenly spaced check-points in Monte Carlo time where every check-point contains $4 \times 30=120$ configurations.
Of those, ground-state (GS) configurations\footnote{
For a $J=\pm 1$ spin-glass, the
ground state is typically highly degenerate even in two
dimensions~\cite{barahona:82b}. Empirically, we find that this
high-degeneracy is also present on the 2D-like Chimera~\cite{katzgraber:14}.
} and minimally-excited (ES) states were picked out for further analysis. 
Our analysis consisted of examining the probability distributions of overlaps between (i) randomly chosen GS configurations and randomly chosen ES configurations
[GS-ES, solid blue curves in Fig.~\ref{fig:Pq} of Extended Data (ED)] and (ii) pairs of randomly chosen GS configurations (GS-GS, dashed red curves in Fig.~\ref{fig:Pq}). 
Two extremal cases were encountered. 
For easy instances (Fig.~\ref{fig:Pq}--top), the probability
  density for GS-GS overlaps is very similar to the probability
  density for GS-ES overlaps.  This is expected in cases where 
  ES states are trivially connected to GS states via one spin-flip. 
Conversely, for hard instances 
  (Fig.~\ref{fig:Pq}--bottom), the probability densities for GS-GS
  overlaps and for GS-ES overlaps differ significantly, indicating that the vast majority of the ES states encountered during the simulations are not trivially connected to typical GS configurations but are rather very distant from them.

The top and bottom panels of Fig.~\ref{fig:Pq} describe only two representative instances, however the above depiction was also found to be valid in the general case, as is confirmed by calculation of the typical $|q|$ for all the instances in each of the hardness groups.\footnote{In order to define a typical
$|q|$, we follow a two-steps procedure. First, we obtain a typical
$|q|$ for each instance by computing the median overlap
  [e.g. for the easy instance in the top panel of Fig.~\ref{fig:Pq}
    $\text{med}(|q|)=0.918$, while for the hard instance in the bottom
    panel $\text{med}(|q|)=0.25$]. Second, we average over all the
instances in a $\tau$-generation, by computing the median over the
instance medians.} As the inset of Fig.~\ref{fig:Pq} shows, the harder instances are, the further away minimally excited states are from ground states. 

\section{The D-Wave Two Chip}\label{sect:DW2}

\subsection{The Chimera}

\begin{figure}
\begin{center}
\includegraphics[angle=90,width=\columnwidth]{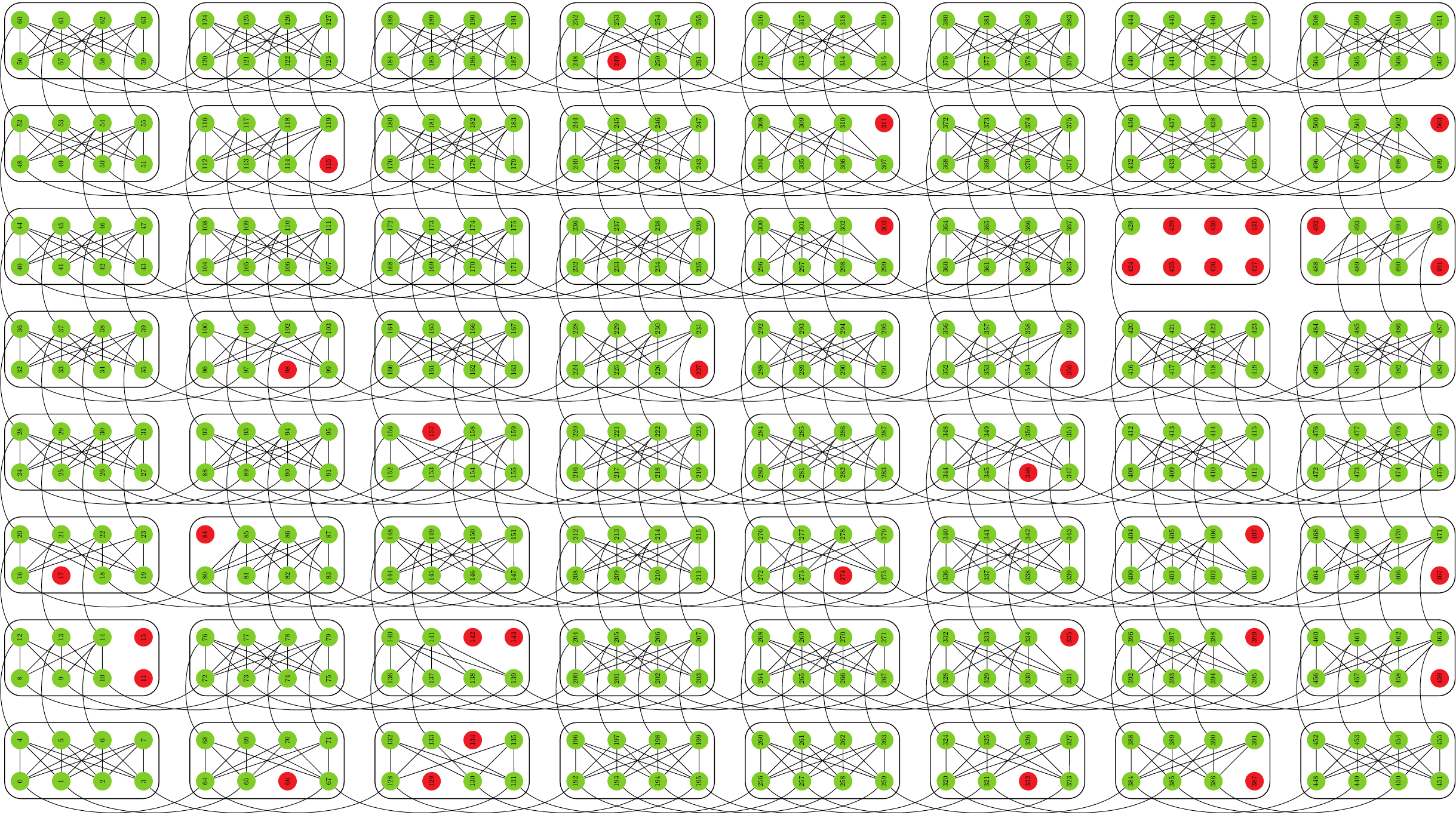}
\caption{{\bf The 476-qubit DW2 device architecture and qubit connectivity.} }
\label{fig:chimera}
\end{center}
\end{figure}

The Chimera graph of the D-Wave Two (DW2) chip used in this study is shown in
Fig.~\ref{fig:chimera}. The chip is an $8 \times 8$ array of unit cells where each unit cell is a balanced $K_{4,4}$
bipartite graph of superconducting flux qubits. In the ideal Chimera graph the
degree of each (internal) vertex is $6$.  On our chip, only a subset of
476 qubits, is functional.  The temperature of the device $\sim 15$mK.

The chip is designed to solve a very specific type of problems, namely, Ising-type optimization
problems where the cost function is that of the Ising Hamiltonian [see Eq.~(1) of the main text]. The Ising spins,
$s_i=\pm 1$ are the variables to be optimized over and the sets
$\{J_{ij}\}$ and $\{h_i\}$ are the programmable parameters of the cost
function.  In addition, $\langle ij\rangle$ denotes a sum over all
active edges of the Chimera graph.\\

\subsection{The D-Wave Two annealing schedule}

\begin{figure}
\begin{center}
\includegraphics[trim=1.55cm 5cm 1.55cm 1.6cm, clip=true,width=\columnwidth]{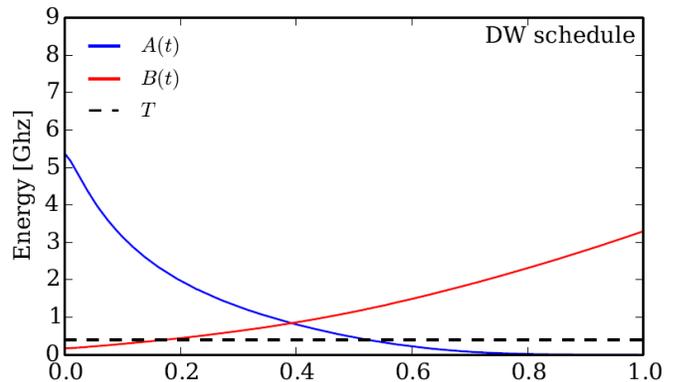}
\caption{{\bf Annealing schedule of the D-Wave chip.} The functions $A(t)$ and $B(t)$ are the amplitudes of the (transverse-field) driver and classical Ising Hamiltonians, respectively. Also shown is the temperature in units of energy ($k_B=1$). }
\label{fig:schedule}
\end{center}
\end{figure}

The DW2 performs the annealing by implementing the time-dependent Hamiltonian
\beq
H(t) = -A(t) \sum_i \sigma_i^x +B(t) H_{\textrm{Ising}} \,,
\eeq
with $t\in [0,t_\ann]$ where the allowed range of annealing times $t_\ann$, due to engineering restrictions, is between $20\mu$s and $20m$s.
The annealing schedules $A(t)$ and $B(t)$ used in the device are shown in Fig.~\ref{fig:schedule}.
There are four annealing lines, and their synchronization becomes harder for faster annealers. 
The filtering of the input control lines introduces some additional distortion in the annealing control.

\subsection{Gauge averaging on the D-Wave device}

Calibration inaccuracies stemming mainly from the digital to analog conversions of the problem parameters, cause the couplings $J_{ij}$ and $h_i$ realized on the DW2 chip to be slightly different from the intended 
 programmed values (with a typical $\sim 5\%$ variation). Therefore, instances encoded on the device will be generally different from the intended instances. Additionally, other, more systematic biases exist which cause spins to prefer one orientation over another regardless of the encoded parameters.
 To neutralize these effects, it is advantageous to perform multiple annealing rounds (or `programming cycles') on the device, where each such cycle corresponds to a different encoding or `gauge' of the same problem instance onto the couplers of the 
 device~\cite{ronnow:14}. To realize these different encodings, we use a gauge freedom in realizing 
 the Ising spin glass: for each qubit we can freely define which of the two qubits states corresponds to $s_i=+1$ and  $s_i=-1$. 
 More formally this corresponds to a gauge transformation that changes spins $s_i\rightarrow  \eta_i s_i$, with $\eta_i=\pm1$ and the couplings as $J_{ij} \rightarrow \eta_i \eta_jJ_{ij}$ and $h_i\rightarrow \eta_ih_i$. The simulations are invariant under such a gauge transformation, but due to calibration errors which break the gauge symmetry, the results returned by the DW2 are not.
 
\subsection{Performance of the DW2 chip as a function of annealing time}\label{sect:ta}

Since the DW2 chip is a putative quantum annealer, it is only natural to ask
how its performance, namely, the typical time-to-solution $\tts$ it yields, depends on annealing time
$t_\ann$. Ideally, the
longer $t_\ann$ is, the better the performance we expect~\cite{kadowaki:98}. However, in practice, decohering interactions with the environment are present which become more pronounced with longer running times of the annealing process. It is therefore plausible to assume that there is an optimal $t_\ann$ for which $\tts$ is shortest~\cite{ronnow:14}.

Analysis of the dependence of success probabilities, or equivalently times to solution, on annealing time is found to be heavily blurred, or masked,  
by fluctuations in the success probability between different programming cycles. These fluctuations, which become more pronounced for harder instances, 
stem from the noisy encoding of the instance parameters already discussed above. Any meaningful analysis of the dependence of success probability on annealing time must therefore successfully average out these effects therefore requires many rounds of anneals.
Unfortunately (see data acquisition in Appendix~\ref{sect:Methods}) the number of annealing
attempts $X$ for a given DW2 programming cycle is proportional to
$1/t_{\text{a}}$, ranging from
$X^{t_{\text{a}}=20\mu\text{s}}=49500$, to
$X^{t_{\text{a}}=20m\text{s}}=49\,.$ As a consequence, the minimal
success probability that can be measured from a
$t_{\text{a}}=20m\text{s}$ programming cycle is just
$p_{\text{min}}^{20m\text{s}}=1/49\approx 0.02$. While this limitation is
not serious for easy instances (i.e., those of the $\tau=10^3$ group), for hard
instances a typical success probability $p$ is much smaller than $0.02$. This problem can
be alleviated by running the DW2 chip on the same instance (and $t_\ann$) multiple number of times (with a different gauge for each cycle, but with $t_\ann$ held fixed), and then averaging the resulting $p$ over programming
cycles. In this way, the minimal success probability that can be
measured is $1/(X N_\text{cycles})$. For $N_\text{cycles}$ in the hundreds typically, typical
numbers for the minimal measurable success probability were
\beq
p_{\text{min}}^{20\mu\text{s}}\approx 6.5\times 10^{-8}\,,\quad
p_{\text{min}}^{200\mu\text{s}}\approx 3.3 \times 10^{-6}\,,\nonumber \\ \nonumber
p_{\text{min}}^{2\text{ms}}\approx 4 \times 10^{-5}\,\text{ and }
p_{\text{min}}^{20\text{ms}}\approx 5.1 \times 10^{-4}\,.
\eeq
However, we have found that the above minimal-probability thresholds are still too
high for our hardest problems, $\tau\geq 10^6$. To overcome this problem, we groups the success
probabilities into annealing-time windows:
$20 \mu\text{s} \leq t_{\text{a}}/10^k  < 60 \mu\text{s}$ and
$60 \mu\text{s} \leq  t_{\text{a}}/10^k < 200 \mu\text{s}$ for $k=0,1,2$
(where in the largest-time interval, $k=2$ above, we also included the
$t_\text{a}=20$ms data).

\begin{figure}
\begin{center}
\includegraphics[angle=270,width=\columnwidth]{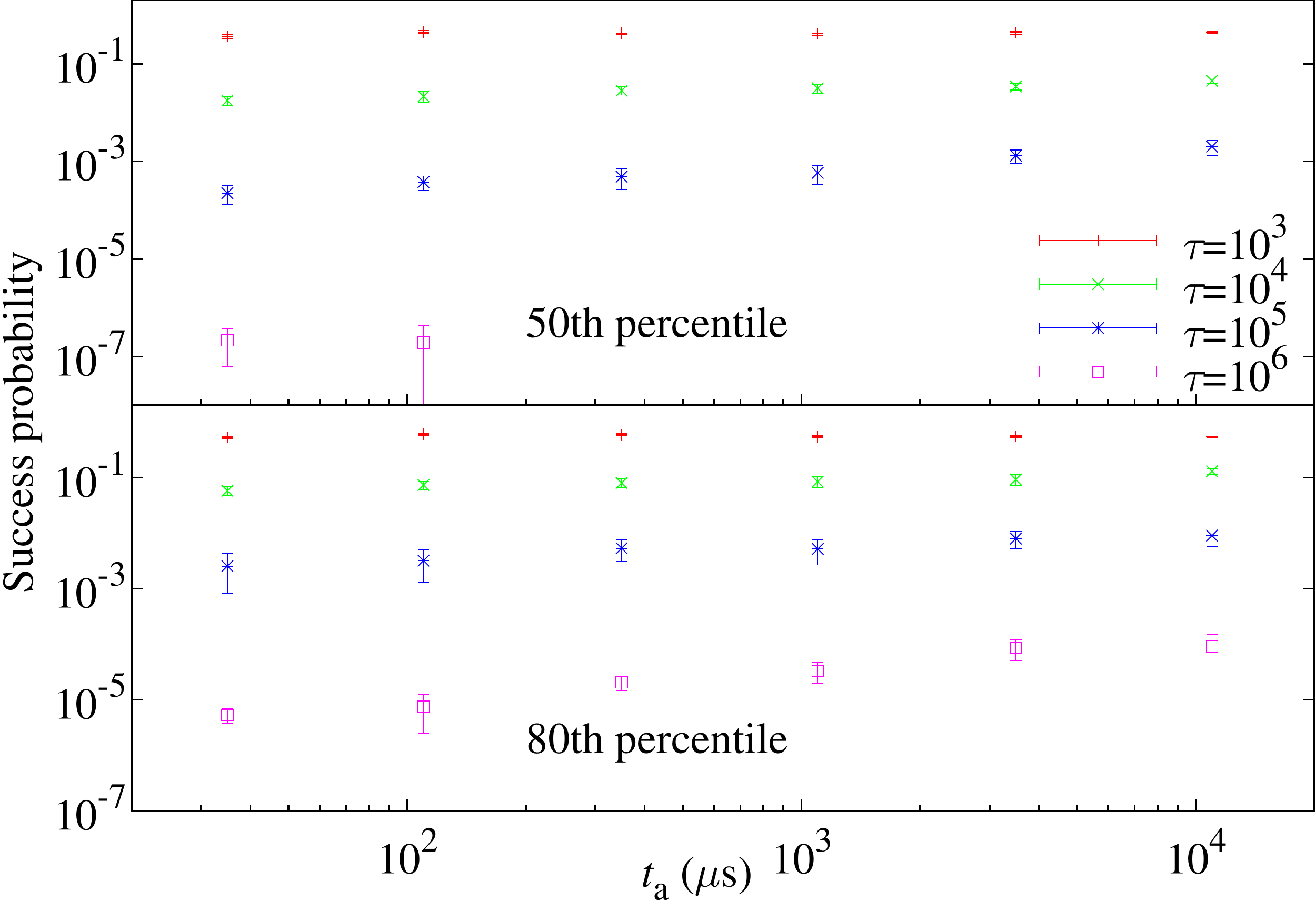}
\caption{{\bf Dependence of success probability on annealing
    time.}  Typical success probability (see main text for details) as a function of  annealing time for the various hardness 
    groups. Shown are the $50$th and $80$th percentiles within each hardness group (top and bottom panels, respectively).}
\label{fig:t_a-dependence}
\end{center}
\end{figure}

At this point, the success probability needs to be averaged over
the problem instances of a given $\tau$-generation. Given the extreme
problem-to-problem fluctuations, percentiles have been calculated. In Fig.~\ref{fig:t_a-dependence}--top we show
the $50$th percentile (i.e., the median) as  a function of probability. Unfortunately, in spite of our
efforts to increase the experimental sensitivity, the measured success
probability yielded zero for more than a half of the instances with
$\tau=10^5$ and $\tau=10^6$.  Hence, we show
Fig.~\ref{fig:t_a-dependence}--bottom the results for the $80$th percentile.

\begin{figure}
\begin{center}
\includegraphics[angle=270,width=\columnwidth]{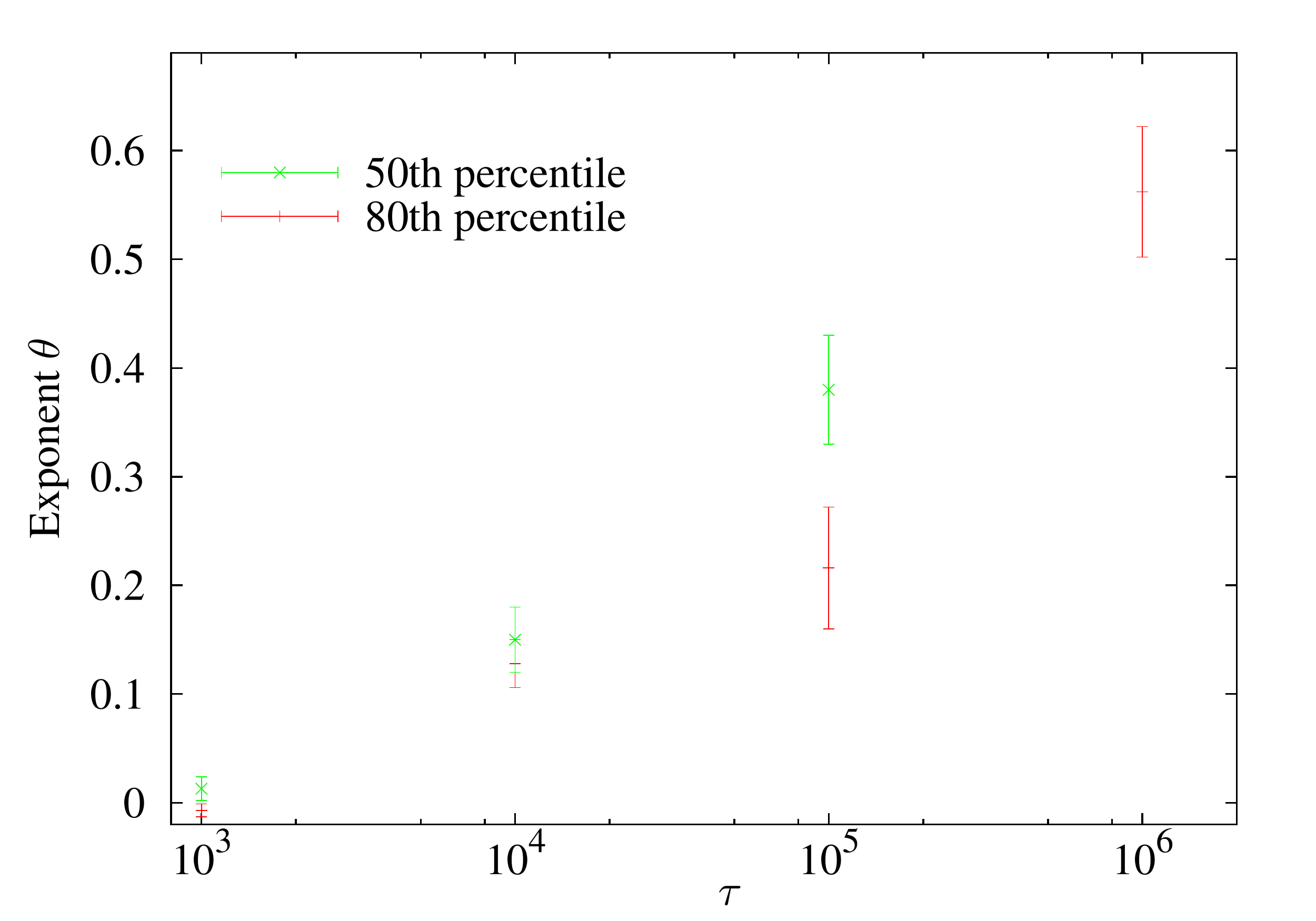}
\caption{{\bf Dependence of success
    probability on annealing time.} The exponent
$\theta$ of Eq.~\eqref{eq:theta-exponent}, as computed from the data shown in
Fig.~\ref{fig:t_a-dependence}, plotted against $\tau$-generation. }
\label{fig:theta-exponent}
\end{center}
\end{figure}

In all cases, we found that a power law description
\beq\label{eq:theta-exponent} p(\tau,t_{\text{a}})\sim
t_{\text{a}}^{\theta(\tau)}\,, \eeq is adequate, although the exponent
$\theta$ depends significantly both on $\tau$ and on the percentile
considered (see Fig~\ref{fig:theta-exponent}).  
The trends are very clear. For easy instances, $p$ barely depends on
$t_{\text{a}}$ (yielding $\theta\approx 0$).  In
fact, the exponent $\theta$ increases with increasing $\tau$, meaning that the harder the instance is, the more it
typically benefits from increasing $t_{\text{a}}$. 

Recalling that time to solution is given by $\tts=t_{\text{a}}/p\sim
t_{\text{a}}^{1-\theta}$, we find that the exponent $\theta$ in
Eq.~\eqref{eq:theta-exponent} is less than 1 for all hardness groups, i.e., that the shorter the annealing time is, the shorter the time-to-solution becomes. 
Since however this trend can not hold all the way down to $t_\ann=0$, these results therefore imply that there exist for each group an optimal annealing time that is however below the shortest-accessible $t_\ann=20\mu$s. Furthermore, the increase of $\theta$ with hardness group can be interpreted as the harder the instances are, the longer the typical optimal annealing time is, consistently with what one would expect from a quantum annealer. It is important to note at this point that the highly-fluctuating success probability, stemming from programming errors, unfortunately does note allow for a more sensitive analysis and that more robust results could be obtained 
if the programing errors leading to $J$-chaos were to be reduced.

\end{document}